\documentclass[twocolumn,prb,showpacs]{revtex4}
\usepackage{graphicx}
\usepackage{dcolumn}
\usepackage{bm}
\usepackage{amsmath}

\begin{document}

\preprint{}
\title{Floquet engineering triplet superconductivity in superconductors with spin-orbit coupling or altermagnetism}
\author{Takehito Yokoyama}
\affiliation{Department of Physics, Institute of Science Tokyo, Tokyo 152-8551,
Japan
}
\date{\today}

\begin{abstract}
We study superconductivitiy under light irradiation based on the Floquet-Magnus expansion in the high-frequency regime. We find that, in spin-singlet superconductors with spin-orbit coupling, triplet superconductivity can be induced in the first-order perturbation for dynamical gap functions  and the second-order perturbation for static gap functions.
We also show that, in unitary triplet superconductors with altermagnetism, nonunitary triplet superconductivity can emerge in the first-order perturbation for dynamical gap function and  in the second-order perturbation for static gap functions.
These results indicate optical generation and control of triplet superconductivity.

\end{abstract}

\maketitle

\section{Introduction}

The ability to manipulate and control the behavior of quantum systems lies at the heart of modern physics\cite{Basov2017,DeLaTorre2021,Disa2021}.
Laser light has been shown to be a versatile tool to control and induce various orders in condensed matter systems, 
e.g., light-induced ferroelectricity\cite{Subedi2015,Mankowsky2017,Subedi2017,Radaelli2018,Nova2019,Li2019,Abalmasov2020,Guo2021,Shin2022,Chen2022,Kwaaitaal2024}, light-induced superconductivity\cite{Cavalleri2017,Knap2016,Babadi2017,Dutreix2017,Fausti2011,Mankowsky2014,Hu2014,Kaiser2014,Fava2024,Mitrano2016,Budden2021,Buzzi2020,Sentef2016,Kennes2017,Claassen2019,Eckhardt2024,Gassner2024}, light-induced magnetism\cite{Kimel2005,Kirilyuk2010,Wang2022,Yang2023,Disa2023,Pitaevskii1960,Ziel1965,Taguchi2011,Fechner2018,RodriguezVega2022},
light-induced multiferroics\cite{Paiva2024}, light-induced chirality\cite{Romao2024,Zeng2024}, and light-induced ferroaxial order\cite{He2023}, enabling remarkable technological advancements and fueling fundamental research. 
In particular, light-driven electron-phonon interactions or symmetry breaking offers a powerful approach to dynamically induce and regulate superconducting phases, paving the way for new quantum states and applications.\cite{Sentef2016,Knap2016,Babadi2017,Dutreix2017,Kennes2017,Eckhardt2024}
In addition to  spin-singlet superconductivity, optical control of spin-triplet superconductivity\cite{Claassen2019} and light-induced switching between singlet and triplet superconducting states\cite{Gassner2024} have been proposed.
Experimentally, evidences for superconducting states in YBCO\cite{Fausti2011,Mankowsky2014,Hu2014,Kaiser2014,Fava2024}, K$_3$C$_{60}$\cite{Mitrano2016,Budden2021} and organic compounds \cite{Buzzi2020} through laser driving in the THz range have been reported.

While static Hamiltonians have traditionally served as the bedrock of quantum mechanics, the burgeoning field of Floquet engineering transcends this paradigm by harnessing time-periodic driving to sculpt novel and exotic quantum phases. This powerful technique opens doors to realizing intriguing phenomena inaccessible in equilibrium systems, offering unprecedented control over material properties and quantum states.\cite{Goldman2014,Bukov2015,Eckardt2017,Oka2019}

Floquet engineering utilizes an external periodic drive, often in the form of electromagnetic fields, to modify the Hamiltonian of a quantum system. This periodic modulation effectively "dresses" the system, leading to the emergence of an effective Floquet Hamiltonian. This new Hamiltonian, unlike its static counterpart, can exhibit entirely new symmetries and topological features, enabling the realization of exotic quantum phenomena\cite{Goldman2014,Bukov2015,Eckardt2017,Oka2019}.

One of the key strengths of Floquet engineering lies in its ability to create novel topological phases of matter. By  tuning the parameters of the periodic driving or inducing effective interactions, researchers can generate Floquet topological phases such as Floquet topological superconductivity that hold immense potential for quantum computing\cite{Benito2014,Ezawa2014,Zhang2015,Takasan2017,Dehghani2021,Kitamura2022}
.


Triplet superconductivity continues to be an area of intense research, especially as new experimental findings and theoretical advancements push the boundaries of our understanding. 
Recent developments have significantly deepened understanding of triplet superconductivity, particularly through the study of candidate materials and the exploration of topological aspects. One of the most noteworthy advancements has been the growing evidence for spin-triplet pairing in unconventional superconductors such as uranium-based compounds, e.g., UTe$_2$, which exhibit robust superconductivity even in the presence of strong magnetic fields.\cite{Aoki2022} These materials have attracted considerable attention due to their potential to support topological superconductivity and host Majorana modes\cite{Alicea2012,Elliott2015}, which are of great interest for quantum computing\cite{Nayak2008}.

Moreover, advancements in spintronics and the study of spin currents in superconductors have opened new pathways for utilizing triplet superconductors in practical applications. The potential for creating devices that exploit the spin degree of freedom in triplet superconductors, combined with their resilience to magnetic fields and Joule heatings, makes them promising candidates for next-generation technologies in quantum computing and beyond.\cite{Linder2015}

In this paper, we study superconductivitiy under illumination of light based on the Floquet-Magnus expansion as shown in Fig. 1. We find that, in spin-singlet superconductors with spin-orbit coupling, triplet superconductivity can be induced in the first-order perturbation for dynamical gap functions  and the second-order perturbation for static gap functions.
We also show that, in unitary triplet superconductors with altermagnetism, nonunitary triplet superconductivity can emerge in the first-order perturbation for dynamical gap function and  in the second-order perturbation for static gap functions.
Modification of triplet superconductivity for triplet spin wave excitation is also discussed.
These results indicate optical generation and control of triplet superconductivity.

This paper is organized as follows: In Sec. II, we introduce the model and Floquet-Magnus expansion for time-periodic Hamiltonian to obtain an engineered effective Hamiltonian for triplet superconductivity.
The results for dynamical gap functions by the first-order perturbation are given in Sec. III A and those for static gap functions by the second-order perturbation are presented in Sec. III B. We summarize our results in Sec. IV.

\section{Formulation}

\begin{figure}[tbp]
\begin{center}
\scalebox{0.8}{
\includegraphics[width=10.0cm,clip]{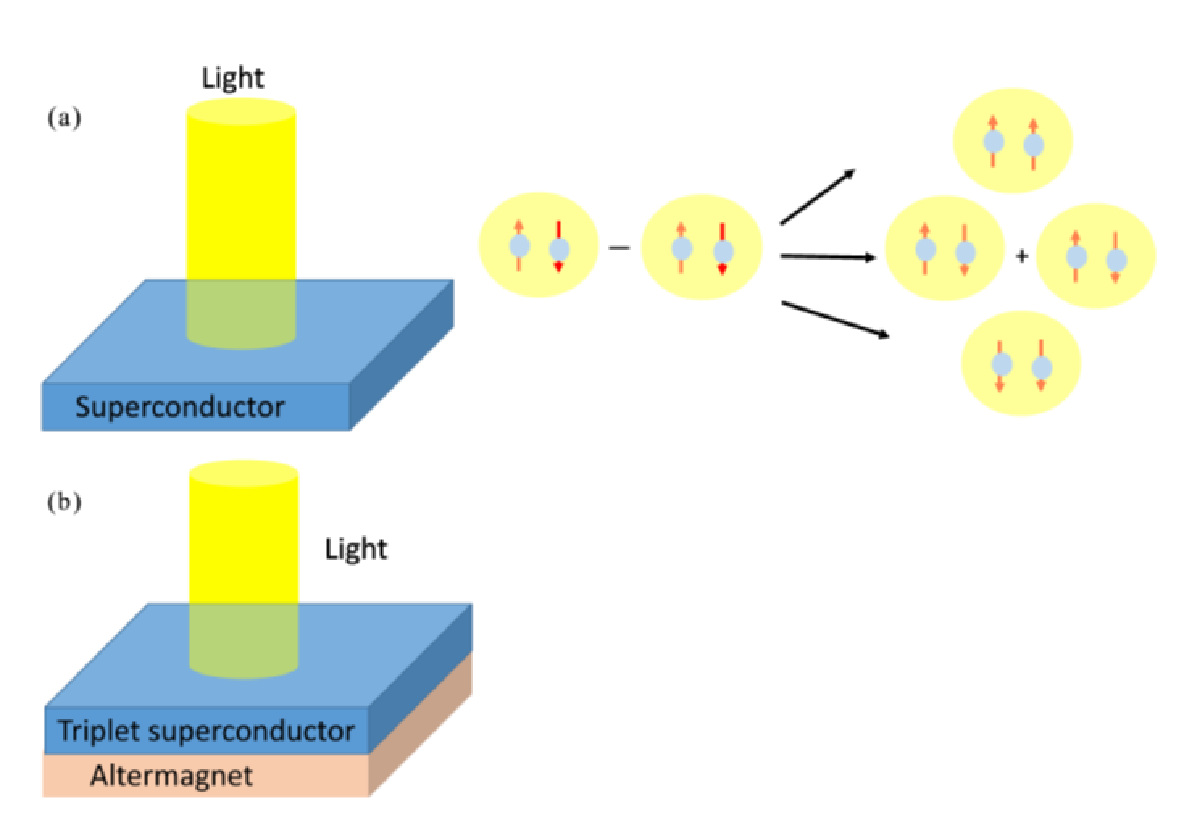}
}
\end{center}
\caption{(Color online) Schematic of the model. (a) Irradiation of the superconductor can engineer triplet superconductivity. For spin-singlet superconductors with spin-orbit coupling, singlet Cooper pairs can be converted into triplet ones. (b) In triplet superconductor/altermagnet junctions, due to the proximity effect, triplet superconductivity and altermagnetism can coexist. }
\label{fig1}
\end{figure}

We consider the Hamiltonian for a superconductor with spin-orbit coupling and exchange field in a matrix form:
\begin{eqnarray}
H = \xi ({\bf{k}}){\tau _3}{\sigma _0} + {\tau _3}{\bf{g}}({\bf{k}}) \cdot {\bm{\sigma }} + {\tau _0}{\bf{h}}({\bf{k}}) \cdot {\bm{\sigma }}\nonumber \\ + {\tau _1}({\Delta _{s}}{\sigma _0} + {\bf{d}}({\bf{k}}) \cdot {\bm{\sigma }}).
\end{eqnarray}
where $ \xi ({\bf{k}})$, ${\bf{g}}({\bf{k}})$, ${\bf{h}}({\bf{k}})$, $\Delta _{s}$, and ${\bf{d}}({\bf{k}})$ denote the kinetic energy, spin-orbit coupling vector, exchange field, spin-singlet gap function and ${\bf{d}}$-vector for spin-triplet gap function, respectively.
$\tau $ and $\sigma$ are Pauli matrices in particle-hole and spin spaces, respectively.
Here, we omit the tensor product. For example, ${\tau _3}{\sigma _0}$ should be interpreted as ${\tau _3} \otimes {\sigma _0}$.
Note that we use the basis such that gap functions for singlet pairing are proportional to $\sigma_0$.\cite{Ivanov2006,Yokoyama2009}
The derivation of this Hamiltonian is given in Appendix A.

Specifically, we consider a superconductor on the square lattice with a Rashba type spin-orbit coupling and $d$-wave exchange field:
\begin{eqnarray}
\xi ({\bf{k}}) =  - 2t\left( {\cos {k_x} + \cos {k_y}} \right) - \mu ,\\ {\bf{g}}({\bf{k}}) = \alpha {\left( { - \sin {k_y},\sin {k_x},0} \right)^T},\\ {\bf{h}}({\bf{k}}) = h\left( {\cos {k_x} - \cos {k_y}} \right){{\bf{e}}_z}
\end{eqnarray}
where ${{\bf{e}}_z}$ denotes the unit vector along $z$-axis. 
The light is described by the vector potential ${\bf{A}}(t)$ which is included by the Peierls  substitutions ${\bf{k}} \to {\bf{k}} +e {\bf{A}}(t)$ and  ${\bf{k}} \to {\bf{k}} -e {\bf{A}}(t)$ in particle and hole spaces, respectively.

We consider the vector potential of the form
\begin{eqnarray}
e{\bf{A}}(t) =  - \frac{e}{\Omega }({E_x}\sin \Omega t,{E_y}\sin (\Omega t + \varphi ),0)^T \nonumber \\ \equiv ({a_x}\sin \Omega t,{a_y}\sin (\Omega t + \varphi ),0)^T.
\end{eqnarray}
$E_x$ and $E_y$ are, respectively, $x$ and $y$ components of the electric field of the light. $\varphi$ describes the phase difference between the $x$ and $y$ components.

We use the Floquet-Magnus expansion for time-periodic Hamiltonian.\cite{Goldman2014,Bukov2015,Eckardt2017,Oka2019}
An effective time-independent Hamiltonian is given as a high frequency expansion with respect to $\Omega^{-1}$.
We consider high frequency expansion up to $\Omega^{-2}$. The effective Hamiltonian then reads
\begin{eqnarray}
{H_F} = {H_0} + \sum\limits_{m = 1}^\infty  {\frac{{\left[ {H_m^\dag ,{H_m}} \right]}}{{m\Omega }}} \nonumber \\ + \sum\limits_{m \ne 0} {\frac{{\left[ {\left[ {H_m^\dag ,{H_0}} \right],{H_m}} \right]}}{{2{m^2}{\Omega ^2}}}}  + \sum\limits_{m \ne 0} {\sum\limits_{n \ne 0,m} {\frac{{\left[ {\left[ {{H_{ - m}},{H_{m - n}}} \right],{H_n}} \right]}}{{3mn{\Omega ^2}}}} }.
\end{eqnarray}
Note that $H_m^\dag  = {H_{ - m}}$ holds due to the hermiticity of the Hamiltonian.

\section{Results}

Let us expand the Hamiltonian and other quantities as
\begin{eqnarray}
O(t) = \sum\limits_{m } {{e^{ - im\Omega t}}} {O_m}
\end{eqnarray}
with $O=H, \xi, {\bf{g}}, {\bf{h}}, \Delta, {\bf{d}}({\bf{k}})$.
By performing the Fourier expansion, we obtain 
\begin{eqnarray}
{\xi _{2m}} =   - 2t\left( {{J_{2m}}({a_x})\cos {k_x} + {e^{ - 2im\varphi }}{J_{2m}}({a_y})\cos {k_y}} \right) - \mu {\delta _{0,m}},\\
{\xi _{2m + 1}} = 2it\left( {{J_{2m + 1}}({a_x})\sin{k_x} + {e^{ - i(2m + 1)\varphi }}{J_{2m + 1}}({a_y})\sin{k_y}} \right)
\end{eqnarray}
\begin{widetext}
\begin{eqnarray}
{\bf{g}}_{2m}({\bf{k}}) = \alpha {\left( {{e^{ - 2im\varphi }}{J_{2m}}({a_y})\sin {k_y}, - {J_{2m}}({a_x})\sin {k_x},0} \right)^T},\\
{{\bf{g}}_{2m + 1}}({\bf{k}}) = i\alpha {\left( {{e^{ - i(2m + 1)\varphi }}{J_{2m + 1}}({a_y})\cos{k_y}, - {J_{2m + 1}}({a_x})\cos{k_x},0} \right)^T},\\
{\bf{h}}_{2m}({\bf{k}}) = h\left( {{J_{2m}}({a_x})\cos {k_x} - {e^{ - 2im\varphi }}{J_{2m}}({a_y})\cos {k_y}} \right){{\bf{e}}_z},\\
{{\bf{h}}_{2m + 1}}({\bf{k}}) =  - ih\left( {{J_{2m + 1}}({a_x})\sin{k_x} - {e^{ - i(2m + 1)\varphi }}{J_{2m + 1}}({a_y})\sin{k_y}} \right){{\bf{e}}_z}.
\end{eqnarray}
\end{widetext}
Here, we have used the Jacobi-Anger expansion 
\begin{equation}
e^{i z \sin \theta} = \sum_{n=-\infty}^{\infty} J_n(z) e^{i n \theta}
\end{equation}
where $J_n$ is the Bessel function.

\subsection{Dynamical gap functions: $\Omega^{-1}$ terms}

Here, we cosider the terms of the order of $\Omega^{-1}$ in the high frequency expansion. To obtain terms relevant to superconductivity in this expansion, the time oscillation of the gap function is necessary. If the gap functions are static, which are proportional to $\tau_{1}$ or $\tau_{2}$, $H_m$ and hence $[H_m^\dag ,{H_m]}$ do not contain terms relevant to superconductivity (proportional to $\tau_{1,2}$). Therefore, dynamical gap functions are required to induce superconductivity in the effective Hamiltonian.
Here, $\Delta_m (m \ne 0)$ in the Fourier expansion
\begin{eqnarray}
\Delta(t) = \sum\limits_{m } {{e^{ - im\Omega t}}} {\Delta_m}
\end{eqnarray}
denotes the dynamical gap function induced by the external drive, while $\Delta_0$ component represents the static gap function. We also use the same notation for ${\bf{d}}$-vector.

One can induce time oscillation of the gap function by directly exciting the amplitude (Higgs) mode of the gap functions or indirectly changing other degree of freedoms such as phonons in the Hamiltionian with irradiation of light. In this paper, we consider two mechanisms: Higgs mode and nonlinear phonon interactions.

(i) We assume that Higgs mode is excited by the light. Since Higgs mode is a scalar excitation of the order parameter, it couples to electromagnetic fields quadratically through ${\bf{A}}^2$. Thus, the Higgs mode is excited when the frequency of the light is equal to the superconducting gap, i.e, $\Omega=\Delta$ and oscillates with the frequency $2\Omega$. \cite{Tsuji2015,Shimano2020}
For details on this mechanism, refer to, e.g., Ref. \cite{Tsuji2015}.

(ii) We assume that infrared-active phonon modes are excited by light and drive Raman phonons with nonlinear interactions that are responsible for the oscillation of the Raman phonons at 2$\Omega$. Consequently, the effective electron-electron interaction becomes dynamical and  the gap functions oscillate at 2$\Omega$.\cite{Knap2016} In this scenario, $\Omega$ and $\Delta$ are independent.
For details on this mechanism, refer to Appendix C in Ref. \cite{Knap2016}.

Therefore, in both cases, the gap functions oscillate with the frequency $2\Omega$. Thus, in this subsection, we consider only even $m$ in the Fourier expansion (i.e., $O_{2m}$ terms).

Then, by using the high frequency expansion, the effective Hamiltonian is obtained as 
\begin{align}
{H_F} = {\xi _0}({\bf{k}}){\tau _3}{\sigma _0} + {\tau _3}\left( {{{\bf{g}}_0}({\bf{k}}) + {{\bf{g}}_a}({\bf{k}})} \right) \cdot {\bm{\sigma }} + {\tau _0}{{\bf{h}}_0}({\bf{k}}) \cdot {\bm{\sigma }}\nonumber \\
 + {\Delta _0}{\tau _1}{\sigma _0} + \left( {{\Delta _a} + {\Delta _b}} \right){\tau _2}{\sigma _0} \nonumber  \\+ {\tau _1}\left( {{{\bf{d}}_0}({\bf{k}}) + {{\bf{d}}_c}({\bf{k}})} \right) \cdot {\bm{\sigma }} + {\tau _2}\left( {{{\bf{d}}_a}({\bf{k}}) + {{\bf{d}}_b}({\bf{k}})} \right) \cdot {\bm{\sigma }}
\end{align}
where 
\begin{eqnarray}
{{\bf{g}}_a}({\bf{k}}) = \frac{2}{\Omega }\sum\limits_{m = 1}^\infty  {\frac{{{\mathop{\rm Im}\nolimits} \left( {{{\bf{g}}_{2m}}({\bf{k}}) \times {{\bf{h}}_{2m}}({\bf{k}})^*} \right)}}{m},}\\ {\Delta _a}({\bf{k}}) = \frac{2}{\Omega }\sum\limits_{m = 1}^\infty  {\frac{{{\mathop{\rm Im}\nolimits} \left( {{\xi _{2m}}({\bf{k}}){\Delta _{2m}}({\bf{k}})^*} \right)}}{m},} \\
{\Delta _b}({\bf{k}}) = \frac{2}{\Omega }\sum\limits_{m = 1}^\infty  {\frac{{{\mathop{\rm Im}\nolimits} \left( {{{\bf{g}}_{2m}}({\bf{k}}) \cdot {{\bf{d}}_{2m}}({\bf{k}})^*} \right)}}{m},}\\ {{\bf{d}}_a}({\bf{k}}) = \frac{2}{\Omega }\sum\limits_{m = 1}^\infty  {\frac{{{\mathop{\rm Im}\nolimits} \left( {{\xi _{2m}}({\bf{k}}){{\bf{d}}_{2m}}({\bf{k}})^*} \right)}}{m},} \\
{{\bf{d}}_b}({\bf{k}}) = \frac{2}{\Omega }\sum\limits_{m = 1}^\infty  {\frac{{{\mathop{\rm Im}\nolimits} \left( {{\Delta _{2m}}({\bf{k}}){{\bf{g}}_{2m}}({\bf{k}})^*} \right)}}{m},}\\ {{\bf{d}}_c}({\bf{k}}) = \frac{2}{\Omega }\sum\limits_{m = 1}^\infty  {\frac{{{\mathop{\rm Im}\nolimits} \left( {{{\bf{h}}_{2m}}({\bf{k}}) \times {{\bf{d}}_{2m}}({\bf{k}})^*} \right)}}{m}}. 
\end{eqnarray}

Let us explain  each term.

(i) ${{\bf{g}}_a}({\bf{k}})$  represents the modification of the Rashba vector by the exchange field.

(ii) ${\Delta _a}({\bf{k}})$ represents the modification of the singlet gap function.

(iii) ${\Delta _b}({\bf{k}})$  describes the conversion of the triplet gap function into singlet one by the spin orbit coupling. 

(iv) ${{\bf{d}}_a}({\bf{k}})$ represents the modification of the triplet gap function.

(v) ${{\bf{d}}_b}({\bf{k}})$ represents the conversion of the singlet gap function into triplet ones by the spin orbit coupling. 

(vi) ${{\bf{d}}_c}({\bf{k}})$  represents the modification of the triplet gap function by the exchange field.

Below, as examples, we consider two specific cases:(1) spin-singlet $s$-wave superconductor with a Rashba type spin-orbit coupling and (2) spin-triplet superconductor with $d$-wave exchange field.

\subsubsection{Spin-singlet $s$-wave superconductor with Rashba type spin-orbit coupling}

For spin-singlet $s$-wave superconductor with a Rashba type spin-orbit coupling, the Hamiltonian is given by 
\begin{align}
H = \xi ({\bf{k}}){\tau _3}{\sigma _0} + {\tau _3}{\bf{g}}({\bf{k}}) \cdot {\bm{\sigma }} + {\tau _1}{\Delta _{s}}.
\end{align}

Thus, we have the following effective Hamiltoinan:
\begin{eqnarray}
{H_F} = {\xi _0}({\bf{k}}){\tau _3}{\sigma _0} + {\tau _3}{{\bf{g}}_0}({\bf{k}}) \cdot {\bm{\sigma }} \nonumber \\+ {\Delta _0}{\tau _1}{\sigma _0} + {\Delta _a}{\tau _2}{\sigma _0} + {\tau _2}{{\bf{d}}_b}({\bf{k}}) \cdot {\bm{\sigma }}.
\end{eqnarray}
We see that an effective triplet gap function described by ${{\bf{d}}_b}({\bf{k}})$ is induced.
This can be expected since the Rashba type spin-orbit coupling breaks inversion symmetry and hence even and odd parity pairings can be mixed.

Assuming ${\Delta _2} = i\Delta '$ with a real $\Delta '$ and taking the $m = 1$ term of the summation over $m$ (validity of the approximation is discussed in Appendix B),
we approximately have 
\begin{eqnarray}
{\Delta _a}({\bf{k}}) \sim \frac{2}{\Omega }{\mathop{\rm Im}\nolimits} \left( {{\xi _2}({\bf{k}}){\Delta _2}({\bf{k}})^*} \right) \\= \frac{{4t\Delta '}}{\Omega }\left( {{J_2}({a_x})\cos {k_x} + \cos2\varphi {J_2}({a_y})\cos {k_y}} \right)\nonumber \\
{{\bf{d}}_b}({\bf{k}}) \sim \frac{{2\alpha \Delta '}}{\Omega }{\left( {\cos2\varphi {J_2}({a_y})\sin {k_y}, - {J_2}({a_x})\sin {k_x},0} \right)^T}.
\end{eqnarray}

Let us estimate the magnitudes of ${\Delta _a}({\bf{k}}) $ and  ${{\bf{d}}_b}({\bf{k}}) $.
For $\Omega=2$meV, $t=1$eV, $\alpha=10$meV, $\Delta'$=0.1meV, $E=10^7$V/m, we have $J_2(0.5) \sim 0.01$, ${\Delta _a}({\bf{k}})  \sim 4t \Delta' J_2(a)/\Omega \sim 1$ meV and ${{\bf{d}}_b}({\bf{k}}) \sim 2\alpha \Delta 'J_2(a)/\Omega \sim 0.01$ meV.

A candidate material to test this prediction is Nb$_{1-x}$Ti$_x$Ni which is an $s$-wave superconductor with spin-orbit coupling. Higgs amplitude mode in this material has been observed.\cite{Matsunaga2013} Also, Nb film deposited on Pt can be used.\cite{Flokstra2023} Spin-orbit coupling itself can induce spin-triplet correlations. However, by comparing  spin-triplet correlations with and without light irradiation, one can experimentally detect the triplet pairings induced by light.

\subsubsection{Spin-triplet superconductor with $d$-wave exchange field}

Superconductivity in altermagnets has been discussed and the coexistence of triplet superconductivity and altermagnetism has been predicted.\cite{Mazin2022,Zhu2023,Brekke2023,Hong2024,Chakraborty2024,Carvalho}
Although there have been no candidate materials that could serve as altermagnetic superconductors with spin-triplet pairing, triplet superconductor/altermagnet junctions, e.g., UTe$_2$/MnTe junction\cite{Aoki2022,Lee2024}, can be used to realize such superconducting states due to the proximity effect as illustrated in Fig. 1(b).

We consider a spin-triplet superconductor with $d$-wave exchange field. Then, the Hamiltonian is given by 
\begin{align}
H = \xi ({\bf{k}}){\tau _3}{\sigma _0} +  {\tau _0}{\bf{h}}({\bf{k}}) \cdot {\bm{\sigma }} + {\tau _1}{\bf{d}}({\bf{k}}) \cdot {\bm{\sigma }}.
\end{align}

The effective Hamiltonian is obtained as
\begin{eqnarray}
{H_F} = {\xi _0}({\bf{k}}){\tau _3}{\sigma _0} + {{\bf{h}}_0}({\bf{k}}) \cdot {\bm{\sigma }} + {\tau _1}\left( {{{\bf{d}}_0}({\bf{k}}) + {{\bf{d}}_c}({\bf{k}})} \right) \cdot {\bm{\sigma }} \nonumber \\+ {\tau _2}{{\bf{d}}_a}({\bf{k}}) \cdot {\bm{\sigma }}.
\end{eqnarray}

The total ${\bf{d}}$-vector in this effective Hamiltonian becomes ${\bf{\tilde d}}({\bf{k}}) = {{\bf{d}}_0}({\bf{k}}) + {{\bf{d}}_c}({\bf{k}}) - i{{\bf{d}}_a}({\bf{k}})$ and hence 
\begin{eqnarray}
{\bf{\tilde d}}({\bf{k}}) \times {\bf{\tilde d}}({\bf{k}})^* = 2i\left( {{{\bf{d}}_0}({\bf{k}}) + {{\bf{d}}_c}({\bf{k}})} \right) \times {{\bf{d}}_a}({\bf{k}}).
\end{eqnarray}
Thus, triplet superconductivity becomes nonunitary for nonzero ${{\bf{d}}_a}$ by light irradiation.

Assuming ${\bf{d}}_2({\bf{k}}) = i{\bf{d}}'$ with a real vector ${\bf{d}}'$, we approximately have 
\begin{eqnarray}
{{\bf{d}}_a}({\bf{k}}) \sim \frac{{4t}}{\Omega }\left( {{J_2}({a_x})\cos {k_x} + \cos2\varphi {J_2}({a_y})\cos {k_y}} \right){\bf{d}}' \\
{{\bf{d}}_c}({\bf{k}}) \sim \frac{{2h}}{\Omega }\left( {{J_{2m}}({a_x})\cos {k_x} - \cos2\varphi {J_2}({a_y})\cos {k_y}} \right){{\bf{e}}_z} \times {\bf{d}}'.
\end{eqnarray}

Let us estimate the magnitudes of ${{\bf{d}}_a}({\bf{k}}) $ and ${{\bf{d}}_c}({\bf{k}})$.
For $\Omega=2$meV, $t=1$eV, $h=10$meV, $d'$=0.01meV, $E=10^7$V/m, we have $J_2(0.5) \sim 0.01$, ${{\bf{d}}_a}({\bf{k}}) \sim 4t d 'J_2(a)/\Omega \sim 0.1$ meV, and ${{\bf{d}}_c}({\bf{k}}) \sim 2h d'J_2(a)/\Omega \sim 0.001$ meV.

\subsection{Static gap functions: $\Omega^{-2}$ terms}

As discussed in the previous section,  for static gap functions, new terms relevant to supeconductivity do not appear in  the order of $\Omega^{-1}$ in the expansion. As for the terms of the order of $\Omega^{-2}$, due to the presence of $H_0$ in the third term in Eq.(6), new terms relevant to supeconductivity emerge even for static gap functions. Thus, we consider $\Omega^{-2}$ terms for static gap functions in the following.

Note that $H_m$ does not contain terms relevant to superconductivity (proportional to $\tau_{1,2}$) and hence the last term in the high frequency expansion does not contain terms relevant to superconductivity. Thus, the last term in the expansion is neglected here. 
Also, in the following, we have calculated terms only relevant to superconductivity, i.e., those with $\Delta _s$ or ${\bf{d}}$.

\subsubsection{Spin-singlet $s$-wave superconductor with Rashba type spin-orbit coupling}

The Fourier coefficients of the Hamiltonian for spin-singlet $s$-wave superconductor with Rashba type spin-orbit coupling read
\begin{align}
{H_{2m}} = {\xi _{2m}}({\bf{k}}){\tau _3}{\sigma _0} + {\tau _3}{{\bf{g}}_{2m}}({\bf{k}}) \cdot {\bm{\sigma }} + {\delta _{0,m}}{\Delta _s}{\tau _1}{\sigma _0} ,\\ {H_{2m + 1}} = {\xi _{2m + 1}}({\bf{k}}){\tau _0}{\sigma _0} + {\tau _0}{{\bf{g}}_{2m+1}}({\bf{k}}) \cdot {\bm{\sigma }}.
\end{align}

Thus, we obtain the effective Hamiltonian:
\begin{widetext}
\begin{align}
{H_F} = {H_0} - \sum\limits_{m = 1}^\infty  {\frac{{{\Delta _s}}}{{{m^2}{\Omega ^2}}}\left[ {\left( {{{\left| {{\xi _{2m}}} \right|}^2} + {{\left| {{{\bf{g}}_{2m}}} \right|}^2}} \right){\tau _1}{\sigma _0} + 2{\tau _1}{\mathop{\rm Re}\nolimits} \left( {\xi _{2m}^*{{\bf{g}}_{2m}}} \right) \cdot {\bm{\sigma }}} \right]}.\end{align}
We find that triplet superconductivity is induced due to the Rashba type spin-orbit coupling. 
The emergent ${\bf{d}}$-vector is appoximately given by
\begin{align}
{\bf{d}}({\bf{k}}) \sim  - \frac{{2{\Delta _s}{\mathop{\rm Re}\nolimits} \left( {\xi _2^*{{\bf{g}}_2}} \right)}}{{{\Omega ^2}}} = \frac{{4{\Delta _s}t\alpha }}{{{\Omega ^2}}} \left( {\begin{array}{*{20}{c}}
{{J_2}({a_y})\sin {k_y}\left( {\cos2\varphi {J_2}({a_x})\cos {k_x} + {J_2}({a_y})\cos{k_y}} \right)}\\
{ - {J_2}({a_x})\sin {k_x}\left( {{J_2}({a_x})\cos {k_x} + \cos2\varphi {J_2}({a_y})\cos{k_y}} \right)}
\end{array}} \right).
\end{align}
The validity of the approximation by only the $m = 1$ term in the summation over $m$ is discussed in Appendix B.

\subsubsection{Spin-triplet superconductor with $d$-wave exchange field}

The Fourier coefficients of the Hamiltonian for spin-triplet superconductor with $d$-wave exchange field  are given by
\begin{align}
{H_{2m}} = {\xi _{2m}}({\bf{k}}){\tau _3}{\sigma _0} + {\tau _0}{{\bf{h}}_{2m}}({\bf{k}}) \cdot {\bm{\sigma }} + {\delta _{0,m}}{\tau _1}{\bf{d}}({\bf{k}}) \cdot {\bm{\sigma }},\; {H_{2m + 1}} = {\xi _{2m + 1}}({\bf{k}}){\tau _0}{\sigma _0} + {\tau _3}{{\bf{h}}_{2m}}({\bf{k}}) \cdot {\bm{\sigma }}.
\end{align}

Thus, the effective Hamiltonian is obtained as 

\begin{align}
{H_F} = {H_0} + {\tau _1}{{\bf{d}}_1}({\bf{k}}) \cdot {\bm{\sigma }} + {\tau _1}{{\bf{d}}_2}({\bf{k}}) \cdot {\bm{\sigma }} + {\tau _1}{{\bf{d}}_3}({\bf{k}}) \cdot {\bm{\sigma }} + {\tau _2}{{\bf{d}}_4}({\bf{k}}) \cdot {\bm{\sigma }},\\
{{\bf{d}}_1}({\bf{k}}) =  - \sum\limits_{m \ne 0}^{} {\frac{{\left( {{{\left| {{\xi _{2m}}} \right|}^2} + {{\left| {{{\bf{h}}_{2m}}} \right|}^2}} \right)}}{{{m^2}{\Omega ^2}}}{\bf{d}}({\bf{k}})}, \\
{{\bf{d}}_2}({\bf{k}}) = \sum\limits_{m \ne 0}^{} {\frac{{\left( {{{\bf{h}}_{2m}}({\bf{k}}) \cdot {\bf{d}}({\bf{k}})} \right)}}{{{m^2}{\Omega ^2}}}{\bf{h}}_{2m}^*({\bf{k}})}, \\
{{\bf{d}}_3}({\bf{k}}) =  - 4\sum\limits_{m}^{} {\frac{{\left( {{\bf{h}}_{2m + 1}^*({\bf{k}}) \cdot {\bf{d}}({\bf{k}})} \right)}}{{{{(2m + 1)}^2}{\Omega ^2}}}{\bf{h}}_{2m + 1}^{}({\bf{k}})}, \\
{{\bf{d}}_4}({\bf{k}}) =  - 2\sum\limits_{m \ne 0}^{} {\frac{{{\mathop{\rm Re}\nolimits} \left( {\xi _{2m}^*{\bf{d}}({\bf{k}}) \times {{\bf{h}}_{2m}}}({\bf{k}}) \right)}}{{{m^2}{\Omega ^2}}}} .
\end{align}
\end{widetext}
Since ${\bf{h}}_{m}^*={\bf{h}}_{-m}$, all these vectors are real.
Let us explain  each term.

(i) ${{\bf{d}}_1}({\bf{k}})$  represents the modification of the ${\bf{d}}$ vector.

(ii) ${{\bf{d}}_2}({\bf{k}})$ and ${{\bf{d}}_3}({\bf{k}})$ represent the modification of the ${\bf{d}}$ vector by the exchange field.

(iii) ${\bf{d}}_4({\bf{k}})$ represent the modification of the ${\bf{d}}$ vector by the exchange field with the phase difference $\pi/2$ compared to ${\bf{d}}$.
We again find that triplet superconductivity becomes nonunitary due to the presence of  ${\bf{d}}_4({\bf{k}})$.

Noting that ${\bf{d}}$ is a real vector and taking the $m=1$ term, 
the ${\bf{d}}_4$ is approximately given by
\begin{align}
{{\bf{d}}_4}({\bf{k}}) \sim  - 4\frac{{{\bf{d}}({\bf{k}}) \times {\mathop{\rm Re}\nolimits} \left( {\xi _{2m}^*{{\bf{h}}_{2m}}} \right)}}{{{\Omega ^2}}} \nonumber \\= \frac{{8th}}{{{\Omega ^2}}} \left( {{J_2}{{({a_x})}^2}{{\cos }^2}{k_x} - {J_2}{{({a_y})}^2}{{\cos }^2}{k_y}} \right){\bf{d}}({\bf{k}}) \times {{\bf{e}}_z}.
\end{align}

Let us estimate the magnitudes of ${\bf{d}}_4({\bf{k}}) $.
For $\Omega=2$meV, $t=1$eV, $h=10$meV, $d$=0.1meV, $E=10^7$V/m, we have 
${{\bf{d}}_4}({\bf{k}}) \sim 8t (J_2(a))^2 d h/\Omega^2 \sim 0.1$ meV.

Here, we provide several comments on the results obtained in this paper:

(i) We have used the high-frequency expansion to derive the effective Hamiltonians. For smaller frequency, the higher order terms in the Floquet-Magnus expansion Eq.(6) should be considered. However, even with the higher order terms, the conclusions in this paper remain qualitatively unchanged.

(ii) We have discussed the emergence of triplet superconductivity by illumination of light. If triplet pairings dominate singlet pairings, topological superconducting phases can be realized. Such phases show an edge state which could be experimentally observed by time-resolved ARPES or time-resolved  STM.

(iii) As for order parameter fluctuations in triplet superconductors, 
in addition to the longitudinal fluctuation of the ${\bf{d}}$ vector, i.e., the amplitude Higgs mode, there are transverse spin wave modes of the ${\bf{d}}$ vector\cite{Kee2000,Chung2012}. Our mechanism also works by excitation of spin wave modes in triplet superconductors, which we discuss in Appendix C.

(iv) If we consider inhomogeneous superconductors such as finite-size superconductors,  we should use a real space formalism. In particular, spatial derivatives of gap functions give new terms relevant to triplet superconductivity in an effective Hamiltonian for inhomogeneous superconductors. We discuss this case in Appendix D.

\section{Summary}
We have investigated superconductivitiy under illumination of light based on the Floquet-Magnus expansion in the high-frequency regime. We have found that, in spin-singlet superconductors with spin-orbit coupling, triplet superconductivity can be induced in the first-order perturbation for dynamical gap functions  and the second-order perturbation for static gap functions.
We have also shown that, in unitary triplet superconductors with altermagnetism, nonunitary triplet superconductivity can emerge in the first-order perturbation for dynamical gap function and  in the second-order perturbation for static gap functions.
These results indicate optical generation and control of triplet superconductivity.

\section{ACKNOWLEDGMENTS}
This work was supported by JSPS KAKENHI Grant No.~JP30578216.

\appendix

\section{Derivation of the Hamiltonian}

\begin{widetext}
Here, we provide the transformation of the basis such that gap functions for singlet pairing are proportional to $\sigma_0$.\cite{Ivanov2006}

The Hamiltonian in the conventional basis is given by
\begin{align}
{H_{}} = \left( {\begin{array}{*{20}{c}}
{\xi ({\bf{k}}) + {\bf{g}}({\bf{k}}) \cdot {\bm{\sigma }} + {\bf{h}}({\bf{k}}) \cdot {\bm{\sigma }}}&{\left( {{\Delta _s}{\sigma _0} + {\bf{d}}({\bf{k}}) \cdot {\bm{\sigma }}} \right){\sigma _2}}\\
{{\sigma _2}\left( {{\Delta _s}{\sigma _0} + {\bf{d}}({\bf{k}}) \cdot {\bm{\sigma }}} \right)}&{ - \xi ({\bf{k}}) + {\bf{g}}({\bf{k}}) \cdot {{\bm{\sigma }}^*} - {\bf{h}}({\bf{k}}) \cdot {{\bm{\sigma }}^*}}
\end{array}} \right).
\end{align}

Now, consider the following unitary matrix
\begin{align}
U = \left( {\begin{array}{*{20}{c}}
{{\sigma _0}}&0\\
0&{{\sigma _2}}
\end{array}} \right).
\end{align}
Under this unitary transformation, the Hamiltonian becomes
\begin{align}
{U^\dag }HU = \left( {\begin{array}{*{20}{c}}
{\xi ({\bf{k}}) + {\bf{g}}({\bf{k}}) \cdot {\bm{\sigma }} + {\bf{h}}({\bf{k}}) \cdot {\bm{\sigma }}}&{{\Delta _s}{\sigma _0} + {\bf{d}}({\bf{k}}) \cdot {\bm{\sigma }}}\\
{{\Delta _s}{\sigma _0} + {\bf{d}}({\bf{k}}) \cdot {\bm{\sigma }}}&{ - \xi ({\bf{k}}) - {\bf{g}}({\bf{k}}) \cdot {\bm{\sigma }} + {\bf{h}}({\bf{k}}) \cdot {\bm{\sigma }}}
\end{array}} \right)
\end{align}
which coincides with Eq.(1).
\end{widetext}

\section{Validity of the approximation}
To check the validity of the approximation by only the $m=1$ term in the summation over $m$, we define
\begin{eqnarray}
F(x,M) = \sum\limits_{m = 1}^M {\frac{{{J_{2m}}(x)}}{m}} ,\; G(x,M) = \sum\limits_{m = 1}^M {\frac{{{J_{2m}}{{(x)}^2}}}{{{m^2}}}}.
\end{eqnarray} 

\begin{figure}[tbp]
\begin{center}
\scalebox{0.8}{
\includegraphics[width=10.0cm,clip]{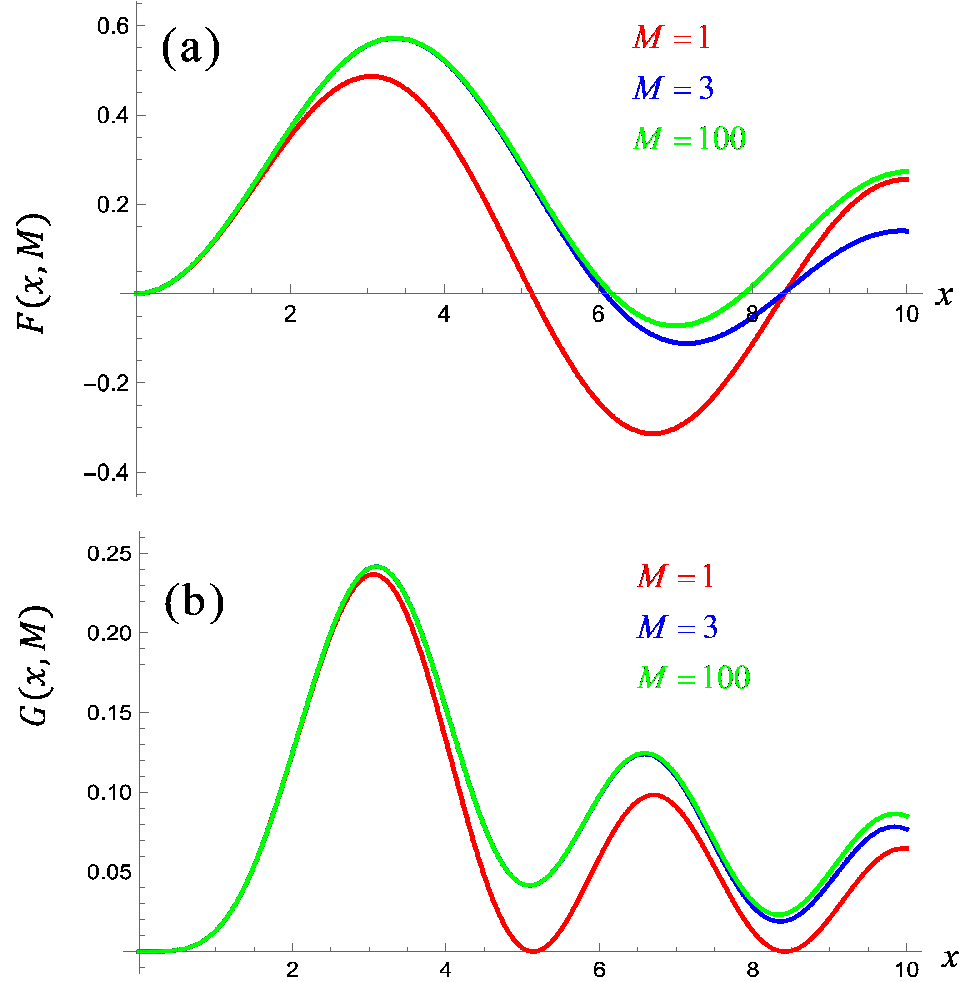}
}
\end{center}
\caption{(Color online) $F(x,M)$ and $G(x,M)$ as a function of $x$. }
\label{fig2}
\end{figure}

$F(x,M)$ is relevant to the $\Omega^{-1}$ terms while $G(x,M)$ is related to the $\Omega^{-2}$ terms.
As seen in Fig.\ref{fig2}, these functions are almost independent of $M$ for $x<2$ in $F(x,M)$ and $x<4$ in $G(x,M)$.
This justifies the approximation by the $m=1$ term of the summation over $m$.

\section{Triplet spin wave}

When an oscillating magnetic field is applied to triplet superconductors, ${\bf{d}}$  vector can precess around the magnetic field according to the Leggett equation (equation of motion for ${\bf{d}}$ vector)\cite{Leggett1974}. Namely, spin wave for ${\bf{d}}$ vector can be excited by oscillating magnetic fields.
Now, let us apply the Floquet approach to this case to see how triplet superconductivity can be engineered. 

The oscillating magnetic field is included as a Zeeman term with 
\begin{align}
{\bf{h}}(t) = (h'\cos\Omega t,h'\sin\Omega t,h)^T.
\end{align}

The Fourier components are obtained as  ${{\bf{h}}_1} = {{\bf{h}}_{ - 1}}^* = h'/2{{\bf{e}}_ + },{{\bf{e}}_ + } = (1,i,0)^T$.

We assume that the ${\bf{d}}$ vector precesess around the $z$-axis:
\begin{align}
{\bf{d}}({\bf{k}}) = d({\bf{k}}){{\bf{e}}_z} + \delta {\bf{d}}
\end{align}
with $\delta {\bf{d}} \cdot {{\bf{e}}_z} = 0$.
The  ${\bf{d}}$ vector oscillates with ${\bf{h}}(t) $. The Fourier component becomes ${{\bf{d}}_1} = d'/2{{\bf{e}}_ + }$.

Then, according to Eq.(28), we have ${{\bf{d}}_a}({\bf{k}}) ={{\bf{d}}_b}({\bf{k}}) =0$ and 
\begin{align}
{{\bf{d}}_c}({\bf{k}}) = \frac{4}{\Omega }{\mathop{\rm Im}\nolimits} \left( {{{\bf{h}}_1} \times {\bf{d}}_1^*} \right) = \frac{{2h'}}{\Omega }{\mathop{\rm Re}\nolimits} \left( {d_{}^{'*}} \right){{\bf{e}}_z}.
\end{align}

We see that the ${\bf{d}}$ vector along the $z$-direction is renormalized by oscillating magnetic field. 

\section{Inhomogeneous superconductors}

We now consider inhomogeneous superconductors such as finite-size superconductors. In such a case, translational symmetry is broken and hence we should use a real space representation. Also, the gap function in general depends on spatial coordinates and thus the commutaton relation between momentum and the gap function is in general nonzero:
\begin{align}
\left[k_i, \Delta(\mathbf{r}) \right] = -i \partial_i \Delta \neq 0
\end{align}
where $\partial_i $ denotes the spatial derivative along the $i$-axis.
Here, we focus on terms with spatial derivative of gap functions pertaining to inhomogeneous superconductors, in the Floquet-Magnus expansion and show how triplet superconductivity can be engineered by inhomogeneity of the gap functions.
We consider two-frequency light irradiation with frequencies $\Omega/2$ and $\Omega$. The light with the frequency $\Omega/2$ is used so that the gap functions oscillate with $\Omega$. The light with the frequency $\Omega$ is included as a vector potential below.

To be specific, we first consider two dimensional $s$-wave superconductors with a Rashba type spin-orbit coupling and a vector potential:
\begin{align}
H = \frac{1}{2m} \left( \mathbf{k}^2 + (e \mathbf{A})^2 \right) \tau_3 \sigma_0 
+ \frac{e}{m} \mathbf{k} \cdot \mathbf{A} \, \tau_0 \sigma_0 \nonumber \\
+ \lambda \tau_3 (k_x \sigma_2 - k_y \sigma_1)
+ e \lambda \tau_0 \left( A_x \sigma_2 - A_y \sigma_1 \right) + \Delta_s(\mathbf{r}) \tau_1 \sigma_0.
\end{align}

We consider the vector potential describing a circularly polarized light:
\begin{align}
\mathbf{A} &= - \frac{E_0}{\Omega} 
\begin{pmatrix}
\sin \Omega t \\
\cos \Omega t \\
0
\end{pmatrix}
 = \frac{E_0}{2\Omega}
\left[
\begin{pmatrix}
i \\
1 \\
0
\end{pmatrix} e^{i \Omega t}
+
\begin{pmatrix}
- i \\
1 \\
0
\end{pmatrix} e^{-i \Omega t}
\right].
\end{align}
For this  vector potential, the Hamiltonian $H$ has terms oscillating at $\pm \Omega$. Note that $\mathbf{A}^2 = \frac{E_0^2}{\Omega^2}$ is a constant.

For time-periodic part of the spin-orbit coupling Hamiltonian
\begin{align}
H_{\mathrm{SO}} = e \lambda \tau_0 (A_x \sigma_2 - A_y \sigma_1),
\end{align}
the commutation relation between $H_{\mathrm{SO}}$ and $H_\Delta = \Delta_s \tau_1 \sigma_0$ is zero:
\begin{align}
[H_{\mathrm{SO}}, H_{\Delta}] = 0.
\end{align}
This means that the first order terms relevant to triplet superconductivity in the Floquet-Magnus expansion vanish.

Let us consider the second order terms in the Floquet-Magnus expansion. The term
\begin{align}
H_A = \frac{e}{m} \mathbf{k} \cdot \mathbf{A} \, \tau_0 \sigma_0
\end{align}
is also time-periodic. Thus, we have
\begin{align}
[H_{A-1}, H_{\Delta 0}] &= - \frac{e}{m} A_{i-1} \partial_i \Delta_s \tau_1 \sigma_0, \\
[[H_{A-1}, H_{\Delta 0}], H_{\mathrm{SO1}}] &= 0.
\end{align}
Note that numbers in the subscripts denote the Fourier coefficients as Eq.(7). We find that the second order terms relevant to triplet superconductivity in the Floquet-Magnus expansion also vanish.

\begin{widetext}
Now, let us turn to two dimensional spin-triplet superconductors with $d$-wave exchange field.
We consider the model Hamiltonian
\begin{align}
H = \frac{1}{2m} \left[ \mathbf{k}^2 +(e\mathbf{A})^2 \right]\tau_3 \sigma_0  + \frac{e}{m} \mathbf{k}\cdot \mathbf{A} \tau_0 \tau_0
+ h \left( k_x^2 +(eA_x)^2 - k_y^2 -(eA_y)^2 \right) \tau_0 \sigma_3 \nonumber \\
+ 2he \left( k_xA_x - k_yA_y \right) \tau_3 \sigma_3
+ \tau_1 \mathbf{d}\cdot \bm{\sigma}
\end{align}
with $\mathbf{d}=(d_x, d_y, 0)$.
We define
\begin{align}
H_h = 2eh \left( k_xA_x - k_yA_y \right) \tau_3 \sigma_3.
\end{align}

Then, we have
\begin{align}
[H_{h-1}, H_{\Delta1}] = \frac{2ieh}{m} \left( A_{x-1} \partial_x d_{x1} \tau_2 \sigma_2 + A_{y-1} \partial_y d_{y1} \tau_2 \sigma_1 \right) ,
\end{align}
\begin{align}
[H_{\Delta-1}, H_{h1}] = -\frac{2ieh}{m}  \left( A_{x1} \partial_x d_{x-1} \tau_2 \sigma_2 + A_{y1} \partial_y d_{y-1} \tau_2 \sigma_1 \right) ,
\end{align}
and hence
\begin{align}
[H_{h-1}, H_{\Delta1}] + [H_{\Delta-1}, H_{h1}] = \frac{ehE_0}{m\Omega}  \left(-( \partial_x d_{x1}+ \partial_x d_{x-1} )\tau_2 \sigma_2 +i(\partial_y d_{y1}-\partial_y d_{y-1} ) \tau_2 \sigma_1 \right).
\end{align}

Therefore, we have the effective Hamiltonian with the first order terms relevant to triplet superconductivity as
\begin{align}
{H_F} = \frac{1}{2m} \left( \mathbf{k}^2 + (\frac{eE_0}{\Omega})^2 \right){\tau _3}{\sigma _0} + {{\mathbf{h}}_0}({\mathbf{k}}) \cdot {\bm{\sigma }} + {\tau _1}{{\mathbf{d}}_0}({\mathbf{k}}) \cdot {\bm{\sigma }} + {\tau _2}\left( {{d_1}{\sigma _1} + {d_2}{\sigma _2}} \right)
\end{align}
with
\begin{align}
{d_1} = i\frac{{eh{E_0}}}{{m{\Omega ^2}}}({\partial _y}{d_{y1}} - {\partial _y}{d_{y - 1}}),\; {d_2} =  - \frac{{eh{E_0}}}{{m{\Omega ^2}}}({\partial _x}{d_{x1}} + {\partial _x}{d_{x - 1}}).
\end{align}
Thus, similar to the case with homogenous superconductors in the main text,  triplet superconductivity becomes nonunitary by illumination of light.

Next, let us consider the second order perturbation. We here consider static gap functions ($d_i=d_{i0}, i=x,y$). By direct calculations, we obtain
\begin{align}
\left[ {\left[ {{H_{h - 1}},{H_\Delta }} \right],{H_{A1}}} \right] =  - \frac{h}{4}{\left( {\frac{{e{E_0}}}{\Omega }} \right)^2}\left( { - {\partial _x} - i{\partial _y}} \right)\left( {i{\partial _x}{d_x}{\tau _2}{\sigma _2} + {\partial _y}{d_y}{\tau _2}{\sigma _1}} \right),
\end{align}
\begin{align}
\left[ {\left[ {{H_{h1}},{H_\Delta }} \right],{H_{A - 1}}} \right] =  - \frac{h}{4}{\left( {\frac{{e{E_0}}}{\Omega }} \right)^2}\left( {{\partial _x} - i{\partial _y}} \right)\left( { - i{\partial _x}{d_x}{\tau _2}{\sigma _2} + {\partial _y}{d_y}{\tau _2}{\sigma _1}} \right).
\end{align}

Thus, we have 
\begin{align}
\left[ {\left[ {{H_{h - 1}},{H_\Delta }} \right],{H_{A1}}} \right] +\left[ {\left[ {{H_{h1}},{H_\Delta }} \right],{H_{A - 1}}} \right] = \frac{{ih}}{2}{\left( {\frac{{e{E_0}}}{\Omega }} \right)^2}\left( {\partial _x^2{d_x}{\tau _2}{\sigma _2} + \partial _y^2{d_y}{\tau _2}{\sigma _1}} \right).
\end{align}

Other commutators in the second order perturbation are calculated in a similar way:
\begin{align}
\left[ {\left[ {{H_{h - 1}},{H_\Delta }} \right],{H_{h1}}} \right] + \left[ {\left[ {{H_{h1}},{H_\Delta }} \right],{H_{h - 1}}} \right] = \frac{2}{m}{\left( {\frac{{eh{E_0}}}{\Omega }} \right)^2}\left( {\partial _x^2{d_x}{\tau _1}{\sigma _1} + \partial _y^2{d_y}{\tau _1}{\sigma _2}} \right)
\end{align}

Hence, we have the effective Hamiltonian for the second order perturbation as
\begin{align}
{H_F} =  \frac{1}{2m} \left( \mathbf{k}^2 + (\frac{eE_0}{\Omega})^2 \right){\tau _3}{\sigma _0} + {{\mathbf{h}}_0}({\mathbf{k}}) \cdot {\bm{\sigma }} + {\tau _1}{\mathbf{d}}({\mathbf{k}}) \cdot {\bm{\sigma }} + {\tau _1}\left( {{d_3}{\sigma _1} + {d_4}{\sigma _2}} \right) + {\tau _2}\left( {{d_5}{\sigma _1} + {d_6}{\sigma _2}} \right)
\end{align}
with
\begin{align}
{d_3} = \frac{1}{m}{\left( {\frac{{eh{E_0}}}{{{\Omega ^2}}}} \right)^2}\partial _x^2{d_x},\; {d_4} = \frac{1}{m}{\left( {\frac{{eh{E_0}}}{{{\Omega ^2}}}} \right)^2}\partial _y^2{d_y},\: {d_5} = \frac{{ih}}{4}{\left( {\frac{{e{E_0}}}{{{\Omega ^2}}}} \right)^2}\partial _y^2{d_y},\; {d_6} = \frac{{ih}}{4}{\left( {\frac{{e{E_0}}}{{{\Omega ^2}}}} \right)^2}\partial _x^2{d_x}.
\end{align}
We again find the effective Hamiltonian similar to Eq.(37). Triplet superconductivity becomes nonunitary by illumination.

\end{widetext}

\end{document}